\def\sample{23.6}

\def\ifb{\ensuremath{{\rm fb}^{-1}}}

\def\BF{\ensuremath{{\cal B}}}

\def\Bs{\ensuremath{B_s^0}}

\def\Bsst{\ensuremath{B_s^{*}}}
\def\Bsstbar{\ensuremath{\overline{B}_s^{*}}}

\def\Bsall{\ensuremath{B_s^{(*)}}}
\def\Bsallbar{\ensuremath{\overline{B}_s^{(*)}}}

\def\BsstBsst{\ensuremath{\Bsst\Bsstbar}}

\def\UfiveS{\ensuremath{\Upsilon(5S)}}
\def\UfourS{\ensuremath{\Upsilon(4S)}}

\def\fBsstBsst{\ensuremath{f_{\Bsst \Bsstbar}}}

\def\NBs{\ensuremath{N_{\Bs}}}

\def\sigmabb{\ensuremath{\sigma_{b\bar{b}}^{\UfiveS}}}

\def\Lint{\ensuremath{L_{\rm int}}}

\def\Bstophig{\ensuremath{\Bs \to \phi \gamma}}

\def\Bstogg{\ensuremath{\Bs \to \gamma \gamma}}

\def\phitoKK{\ensuremath{\phi \to K^+ K^-}}

\def\BztoKstg{\ensuremath{B^{0}\to{} K^*(892)^{0} \gamma}}
\def\BptoKstg{\ensuremath{B^{+}\to{} K^*(892)^{+} \gamma}}
\def\BtoKstg{\ensuremath{B \to{} K^*(892) \gamma}}

\def\BtoXsg{\ensuremath{B \to X_s \gamma}}

\def\gev{\ensuremath{\rm GeV}}

\def\mbc{\ensuremath{M_{\mathrm{bc}}}}
\def\deltae{\ensuremath{{\Delta}E}}

\def\thhel{\ensuremath{\theta_{\rm hel}}}
\def\costhhel{\ensuremath{\cos{\thhel}}}

\def\Ebeam{\ensuremath{E_{\rm beam}^{\rm CM}}}

\def\yieldBsstBsst{\ensuremath{S_{\Bsst\Bsstbar}}}

\def\yieldBsstBsstBstophig{\ensuremath{18 {^{+6}_{-5}}}}

\def\bfBstophigsu{\ensuremath{57 { ^{+18}_{-15}  } { ^{+12}_{-17}} }}
\def\bfBstophigss{\ensuremath{(57 { ^{+18}_{-15}(\rm stat) } { ^{+12}_{-17} (\rm syst) }} ) \times 10^{-6} }

\def\systsignifbfBstophig{\ensuremath{5.5}}

\def\effBstophig{\ensuremath{24.7 \pm 0.2}}

\def\yieldBsstBsstBstogg{\ensuremath{-6.8 {^{+2.4}_{-1.9} } }}

\def\limitbfBstogg{\ensuremath{8.6 \times 10^{-6}}}
\def\limitbfBstoggsu{\ensuremath{8.6}}

\def\effBstogg{\ensuremath{17.8 \pm 0.2}}

\documentclass[a4paper]{jpconf}
\usepackage{graphicx}
\usepackage{url}

\begin{document}


\vspace{-10mm}

\title{Radiative penguin \boldmath $B_s$ decays at Belle}
\author{Jean Wicht for the Belle Collaboration}

\address{\'Ecole Polytechnique F\'ed\'erale de Lausanne (EPFL), Lausanne, Switzerland}

\ead{jean.wicht@epfl.ch}

\begin{abstract}
We report searches for the radiative penguin decays \Bstophig\ and \Bstogg\ based on a \sample\ \ifb\ data 
sample collected with the Belle 
detector at the KEKB $e^+e^-$ energy-asymmetric collider operating at the \UfiveS\ resonance. We obtain the first observation of a radiative penguin decay of the \Bs\ meson in the \Bstophig\ mode and we measure $\BF(\Bstophig) = \bfBstophigss$. No significant \Bstogg\ signal is observed and we set an upper limit at 90\% confidence level of $\BF(\Bstogg) < \limitbfBstogg$. These results are preliminary.
\end{abstract}

\section{Introduction}
\begin{figure}
\centering
\begin{tabular}{ccc}
   \includegraphics[width=0.30\textwidth]{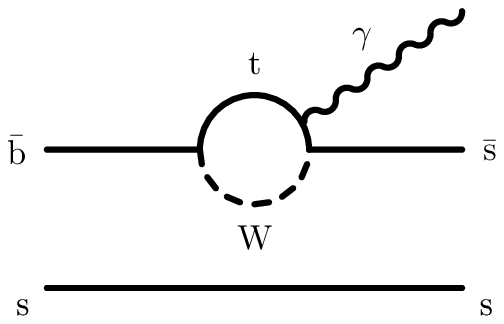} &
   \hspace{1cm} &
   \includegraphics[width=0.30\textwidth]{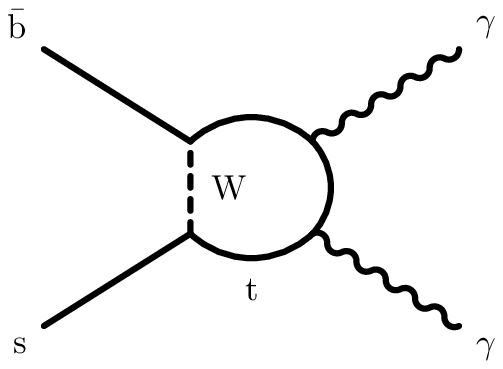} \\
\end{tabular}
\caption{Diagrams of the dominant processes for the \Bstophig\ (left) and \Bstogg\ (right) decays.}
\label{figure:feynman}
\end{figure}

The \Bstophig\ mode is a radiative penguin decay characterized by the $b \to s \gamma$ quark transition (Fig.~\ref{figure:feynman} left); it is the strange counterpart of the \BtoKstg\ decay whose observation by CLEO in 1993~\cite{b2kstg-cleo} proved the existence of penguin processes. In the Standard Model (SM), the branching fraction of \Bstophig\ is predicted to be $(39.4 \pm 11.9) \times 10^{-6}$~\cite{bs2phigam-sm}. The \Bstogg\ mode is a penguin annihiliation decay (Fig.~\ref{figure:feynman} right) and its branching fraction has been calculated in the SM to be in the range $(0.5-1.0) \times 10^{-6}$~\cite{bs2gamgam-sm1,bs2gamgam-sm2,bs2gamgam-sm3}. \Bstophig\ and \Bstogg\ have not been observed yet and the most stringent limit at 90\% confidence level (CL) on their branching fractions are respectively $1.2 \times 10^{-4}$~\cite{bs2phigam-cdf} and $53 \times 10^{-6}$~\cite{u5s-excl}. 

The study of radiative penguin decays is a good tool to search for physics beyond the SM. A strong constraint on the \Bstophig\ branching fraction is generally expected due to the good agreement between SM expectations and experimental results in $b \to s \gamma$ rates such as in the \BptoKstg\ and \BztoKstg\ decays~\cite{b2kstg-th,PDG2007} or the inclusive $B \to X_s \gamma$ decay~\cite{b2xsg-th,hfag}. The \Bstogg\ decay is constrained in a similar way~\cite{bs2gamgam-xsg} but New Physics (NP) scenarios such as supersymmetry with broken R-parity~\cite{bs2gamgam-supersym}, a fourth quark generation~\cite{bs2gamgam-4thquark} or the two Higgs doublet model with flavor changing neutral currents~\cite{bs2gamgam-2higgsdoublet}, can increase the \Bstogg\ branching fraction up to one order of magnitude and still provide a small contribution to \BtoXsg. 

\section{Data sample and analysis}

In this study, we use a data sample with an integrated luminosity (\Lint) of \sample\ \ifb\ that were 
collected with the Belle detector~\cite{Belle} at the KEKB asymmetric-energy $e^+e^-$ 
(3.6 on 8.2~\gev) collider~\cite{KEKB} operating at the $\UfiveS$ resonance. The variety of hadronic events at the \UfiveS\ resonance is richer than at \UfourS. $B^+$, $B^0$ and \Bs\ mesons are produced through the decay of the \UfiveS. The \Bs\ mesons are mostly produced in the $\UfiveS \to \BsstBsst$ decay channel, with the subsequent decays of the excited \Bsst\ states to the ground states with the emission of a slow photon. Therefore, we search for \Bstophig\ and \Bstogg\ in \BsstBsst\ events. The $b\bar{b}$ production cross-section at \UfiveS\ has been measured to be $\sigmabb = (0.302 \pm 0.015)$ nb~\cite{u5s-incl}, the fraction of $\Bsall\Bsallbar$ events in $b\bar{b}$ events to be $f_s = N(\Bsall\Bsallbar)/N(b\bar{b}) = (19.5^{+3.0}_{-2.2})\%$~\cite{PDG2007} and the fraction of $\Bsst\Bsstbar$ events among $\Bsall\Bsallbar$ events to be $\fBsstBsst = (93^{+7}_{-9})\%$~\cite{u5s-excl}. 

We reconstruct $\phi$ mesons in the decay mode \phitoKK. \Bs\ mesons are selected by means of the beam-energy constrained mass $\mbc = \sqrt{(\Ebeam)^2-(p_{\Bs}^{\rm CM})^2}$ and the energy difference $\deltae = E_{\Bs}^{\rm CM} - \Ebeam$ where $p_{\Bs}^{\rm CM}$ and $E_{\Bs}^{\rm CM}$ are the momentum and the energy of the \Bs\ meson, all variables being evaluated in the center-of-mass (CM) frame. \Bsst\ mesons are not fully reconstructed due to the low energy of the photon from the \Bsst\ decay. The main background is due to continuum events coming from light-quark pair production ($e^+e^- \to u\bar{u},d\bar{d},c\bar{c},s\bar{s}$). This background is rejected using a Fisher discriminant based on modified Fox-Wolfram moments describing event topology and a veto of $\pi^0$ and $\eta$ mesons decaying to two photons.
For the \Bstophig\ (\Bstogg) mode, we perform a three-dimensional (two-dimensional) unbinned maximum likelihood fit to \mbc, \deltae\ and \costhhel\ (\mbc\ and \deltae). The helicity angle \thhel\ is the angle between the \Bs\ and the $K^+$ evaluated in the $\phi$ rest frame.

Both fits have four free fit variables: the branching fraction, the continuum background normalization, the \mbc\ continuum shape parameter and the continuum \deltae\ slope. The signal yield is defined as $\yieldBsstBsst = \BF \times \epsilon \times \NBs \times \fBsstBsst$, where $\BF$ is $\BF(\Bstophig) \times \BF(\phitoKK)$ for the \Bstophig\ mode and $\BF(\Bstophig)$ for the \Bstogg\ mode, $\epsilon$ is the MC signal efficiency listed in Table~\ref{table:results} and $\NBs = 2 \times \Lint \times \sigmabb \times f_s = (2.8 {^{+0.5}_{-0.3}}) \times 10^6$ is the number of \Bs\ mesons.

\section{Results and conclusion}

We observe \yieldBsstBsstBstophig\ signal events in the \Bstophig\ mode and measure $\BF(\Bstophig) = \bfBstophigss$ with a significance of 5.5 $\sigma$.  Results are reported in Table~\ref{table:results} and fit projections are shown in Fig.~\ref{figure:data}. This is the first observation of a radiative penguin decay of the \Bs\ meson. The measured branching fraction is in agreement with SM expectation and with the measurement of the branching fractions of the \BptoKstg\ and \BztoKstg\ decays~\cite{PDG2007}. 

\begin{figure}
\centering
 \begin{tabular}{ccc}
   \includegraphics[width=0.30\textwidth]{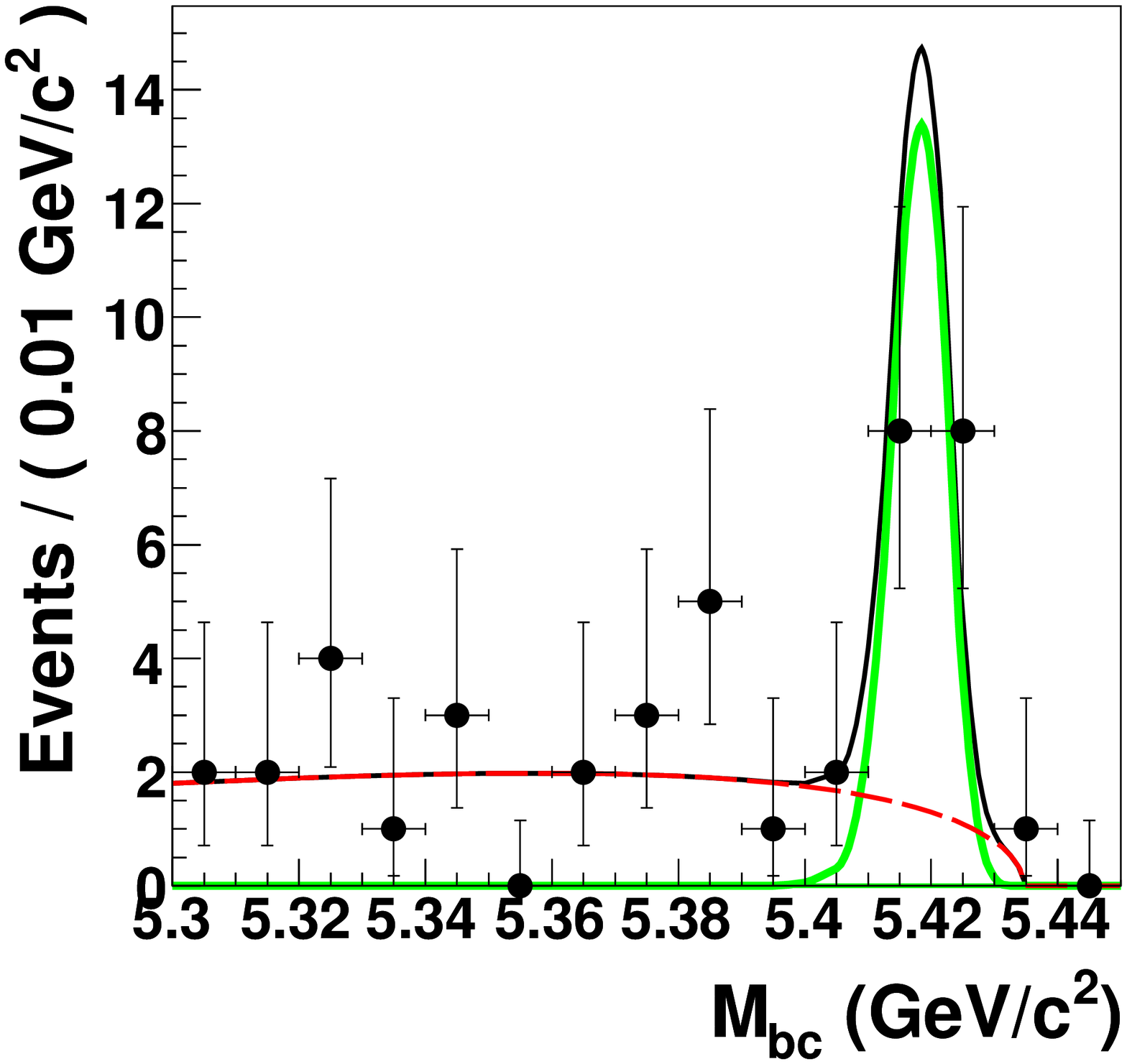} & 
   \includegraphics[width=0.30\textwidth]{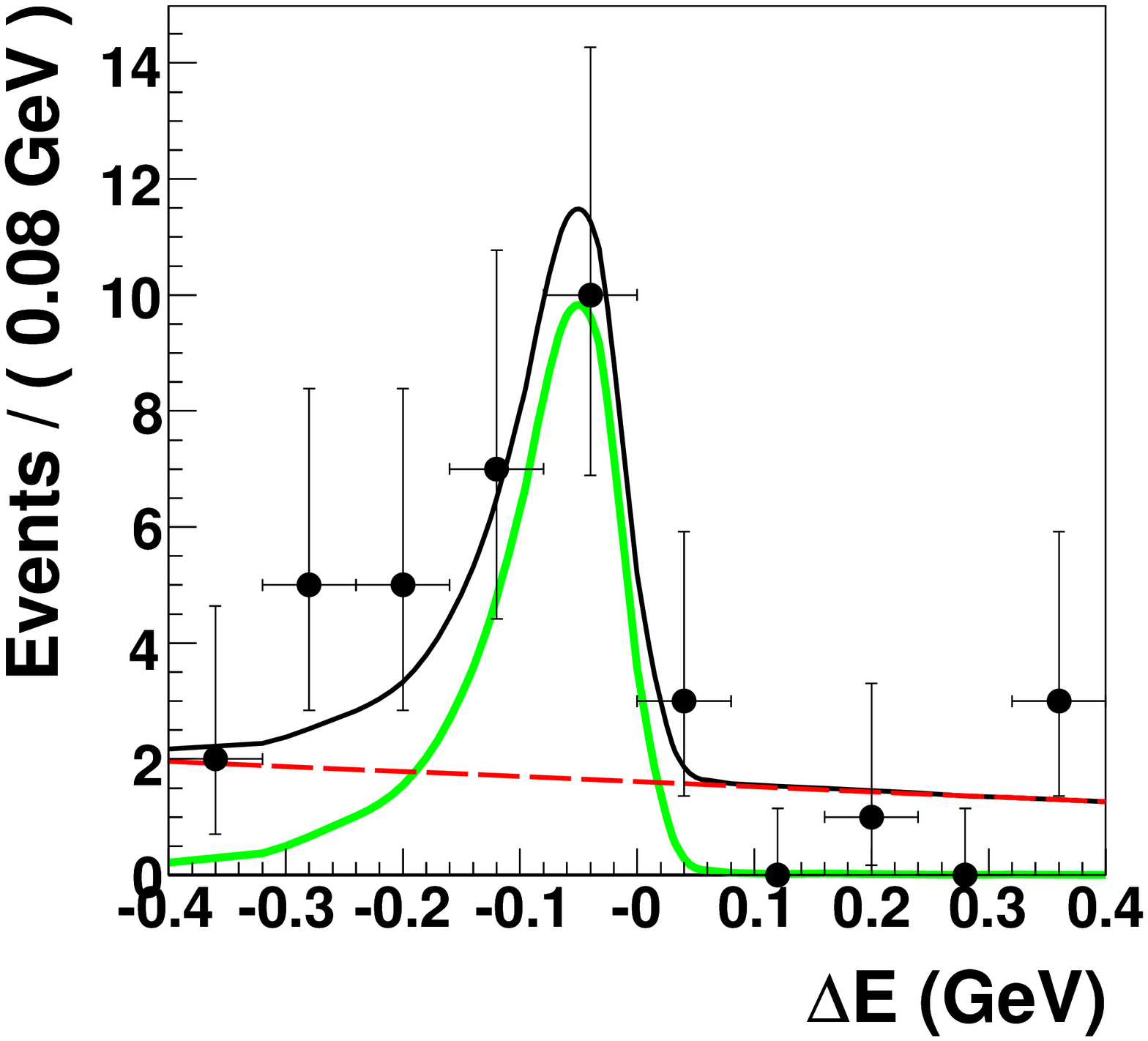} & 
   \includegraphics[width=0.30\textwidth]{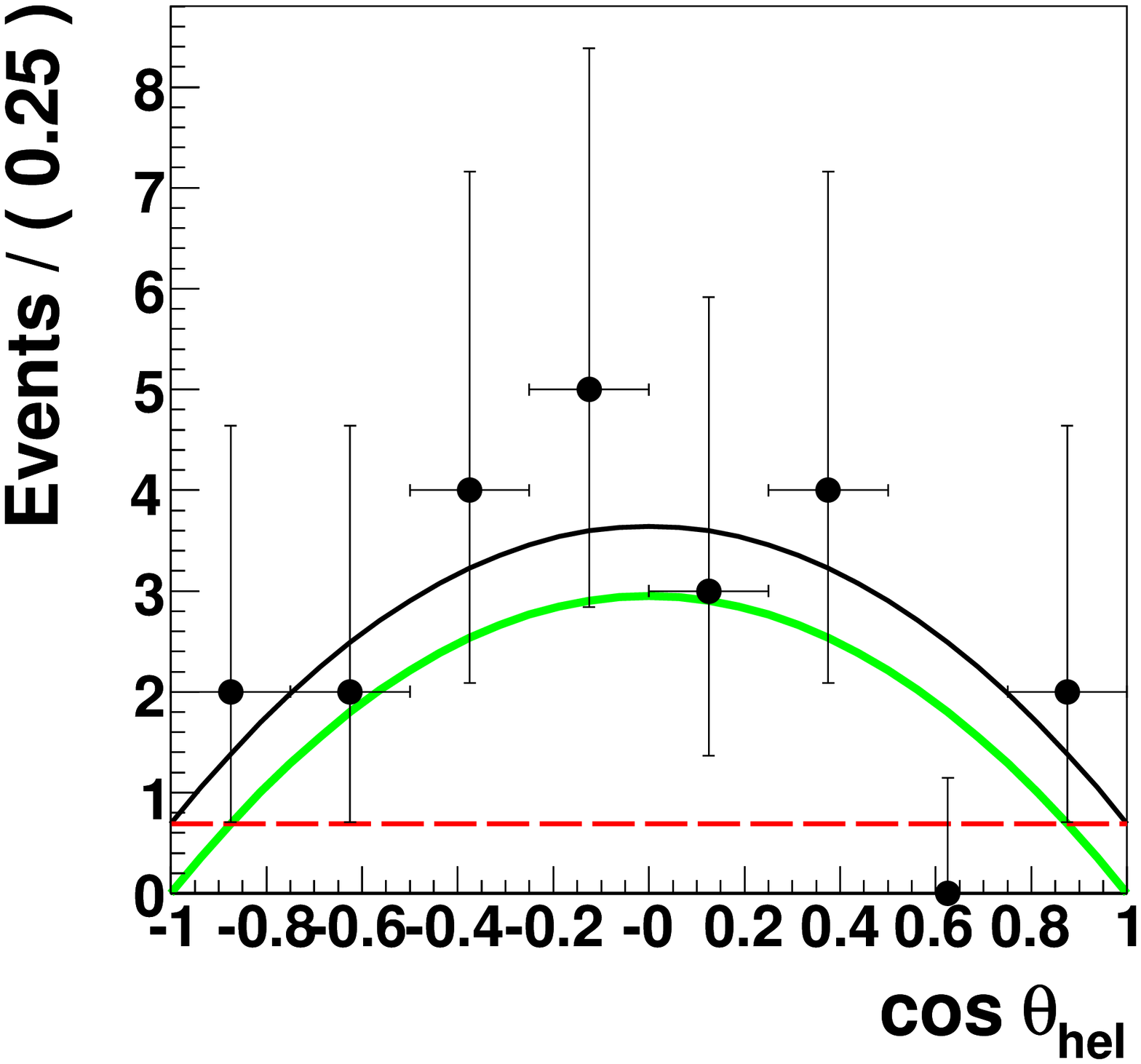} \\   
   \includegraphics[width=0.30\textwidth]{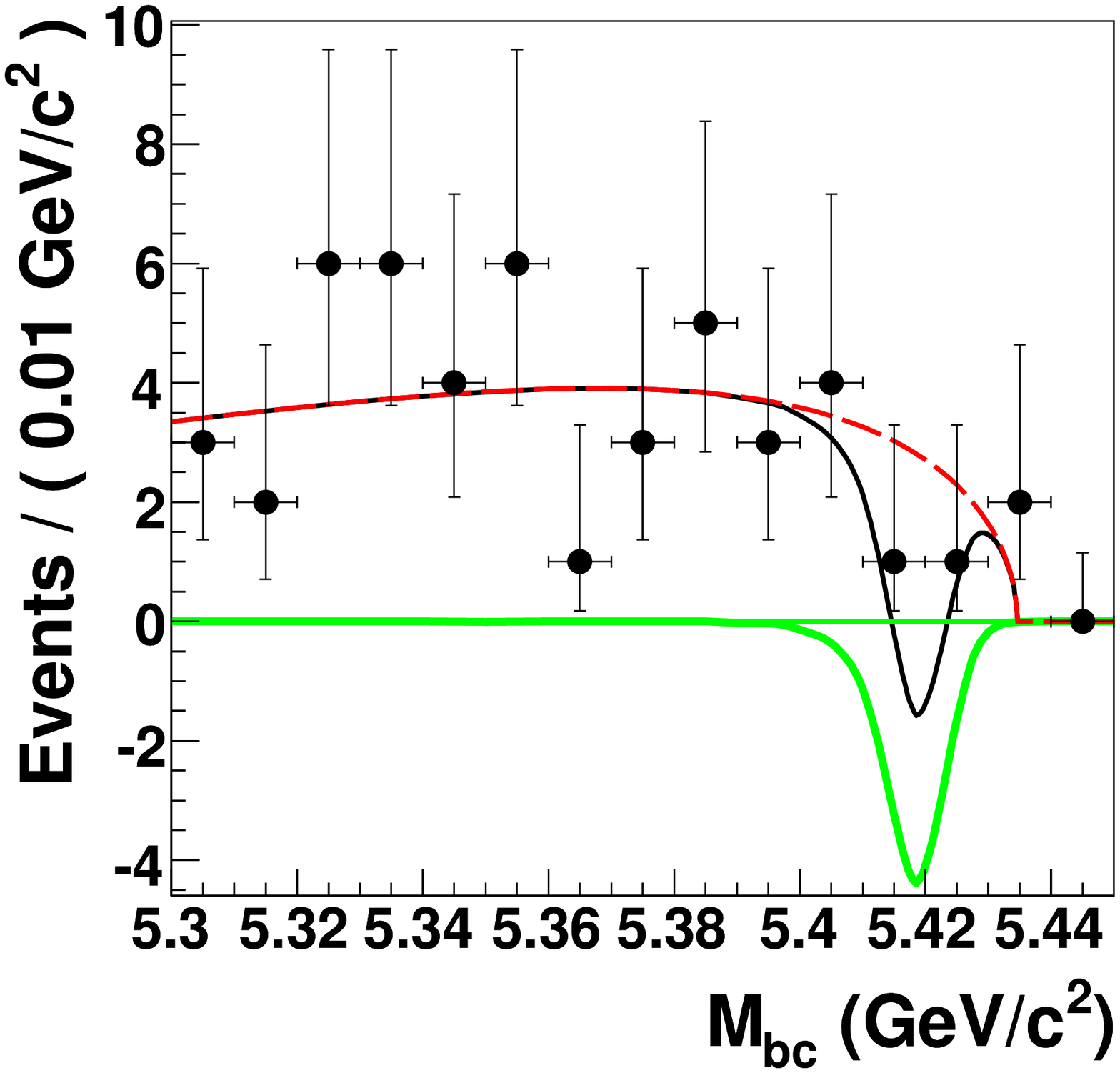} & 
   \includegraphics[width=0.30\textwidth]{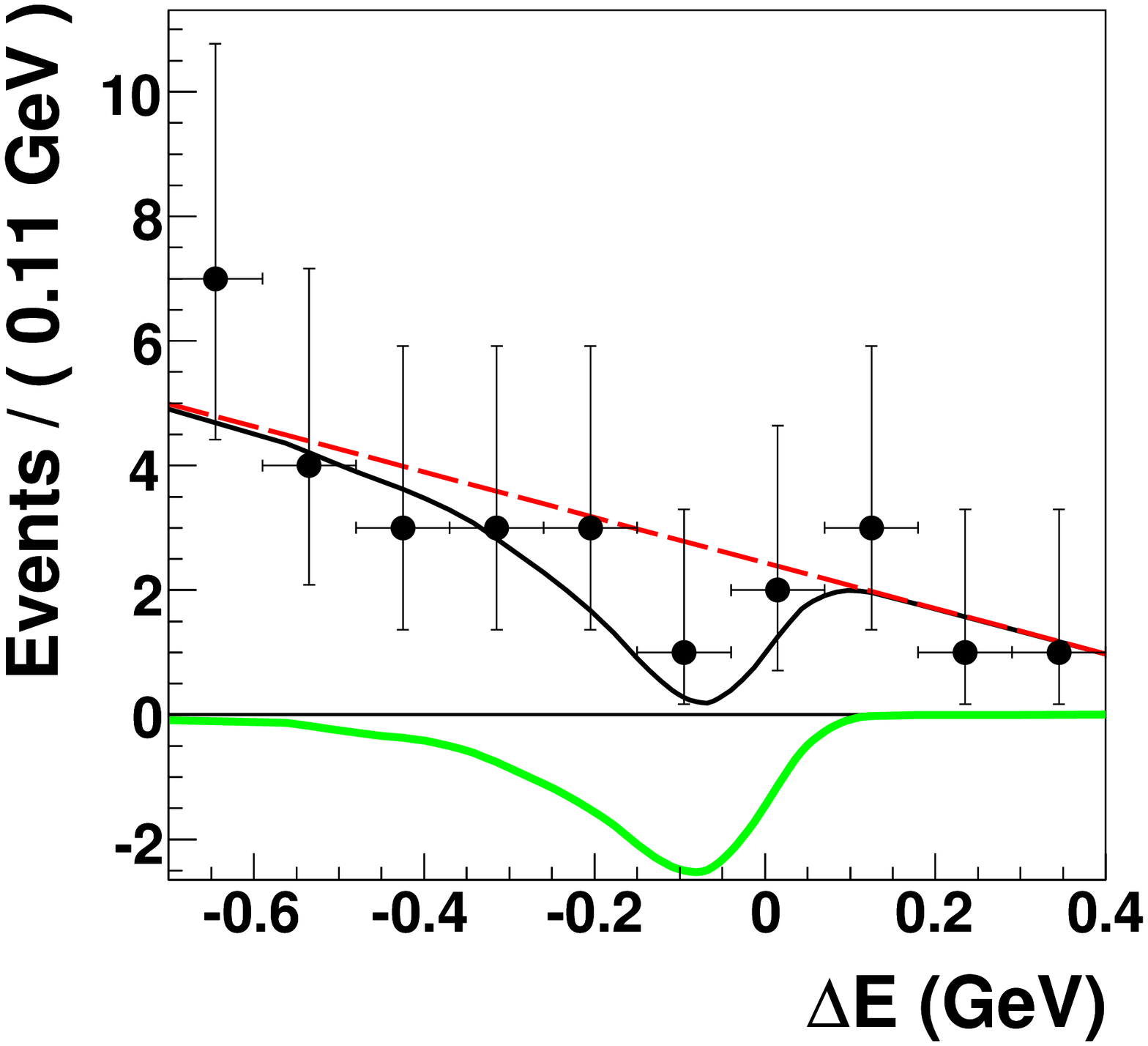} & \\
 \end{tabular}
\caption{\mbc, \deltae\ and \costhhel\ projections together with fit results for the \Bstophig\ mode (top) and the \Bstogg\ mode (bottom). The points with error bars represent data, the thin solid curves are
 the fit functions, the thick solid curves are the signal function and the dashed lines show the continuum contribution.}
\label{figure:data}
\end{figure}

\begin{table}
\centering
\caption{Efficiencies ($\epsilon$), signal yields, branching fractions and significances ($\mathcal{S}$) obtained from the fit. The first uncertainty is statistical, the second one is systematic. The upper limit is calculated at 90\% CL.}
\begin{tabular*}{\textwidth}{@{\extracolsep{\fill}}lcccr}\hline
Mode      & $\epsilon$ (\%) & \yieldBsstBsst         & \BF\ ($10^{-6}$)     & $\mathcal{S}$          \\ \hline 
\Bstophig & \effBstophig    & \yieldBsstBsstBstophig & \bfBstophigsu         & \systsignifbfBstophig     \\
\Bstogg   & \effBstogg      & \yieldBsstBsstBstogg   & $< \limitbfBstoggsu $  & --   \\ \hline
\end{tabular*}
\label{table:results}
\end{table}

We see no significant signal in the \Bstogg\ mode and we extract an upper limit at 90\% CL of $\BF(\Bstogg) < \limitbfBstogg$. This limit, obtained with an integrated luminosity of \sample\ \ifb, is significantly more stringent than the published one~\cite{u5s-excl}, but still above the current NP predictions. However, it is only one order of magnitude larger than the SM prediction leaving good hope for a Super $B$-factory~\cite{superbelle} to observe this decay in the future.

\end{document}